\documentclass[traditabstract]{aa} 
\usepackage{graphicx,lscape,longtable,natbib}

\usepackage{txfonts}
\newcommand{\ltsima}{$\; \buildrel < \over \sim \;$}
\newcommand{\simlt}{\lower.5ex\hbox{\ltsima}} 
\newcommand{\gtsima}{$\; \buildrel > \over \sim \;$}
\newcommand{\simgt}{\lower.5ex\hbox{\gtsima}} 

\newcommand{\xmm}{{\emph{XMM-Newton}}}

\newcommand{\lum}{erg~s$^{-1}$}
\newcommand{\flux}{{erg~cm$^{-2}$~s$^{-1}$ }}

\newcommand{\mum}{\:\mu\mbox{\scriptsize m}}

\newcommand{\sorg}{IRAS~19254-7245}

\begin{document}

\title{A Suzaku observation of the ULIRG IRAS19254-7245: disclosing the AGN 
component.}

\author{V. Braito
\inst{1,2} 
 \and 
 J.N. Reeves
 \inst{2,3}
 \and   
 R. Della Ceca
 \inst{4}
 \and 
 A. Ptak
 \inst{2,5}
 \and  G. Risaliti
 \inst{6,7}
 \and  T. Yaqoob,
 \inst{2,5}}   
\institute{
X-Ray Astronomy Group, Department of Physics and Astronomy, Leicester University, Leicester LE1 7RH, UK 
\email bv67@star.le.ac.uk
\and
Department of Physics and Astronomy, Johns Hopkins University, Baltimore, MD 21218.
\and 
Astrophysics Group, School of Physical and Geographical Sciences, Keele University, Keele, Staffordshire ST5 5BG
\and
INAF-Osservatorio Astronomico di Brera, via Brera 28, I-20121 Milan, Italy.
\and 
Astrophysics Science Division, Code 662, NASA/Goddard Space Flight Center, Greenbelt, MD 20771, USA
\and
INAF-Osservatorio Astrofisico di Arcetri, Largo E. Fermi 5, I-50125 Florence, Italy.
\and 
Harvard-Smithsonian Center for Astrophysics, 60 Garden Street, Cambridge, MA 02138.}

\abstract
  {We discuss here a long Suzaku observation of IRAS 19254-7245  (also known as
the Superantennae), one of the brightest and well studied Ultra Luminous Infrared
Galaxies in the local Universe. \\  This long observation provided the first
detection of \sorg\ above 10 keV,  and  measured a 15--30 keV flux  of  $\sim 5\times
10^{-12}$\flux.
The detection above 10 keV has allowed us to
unveil, for the first time, the  intrinsic luminosity  of the AGN hosted in
\sorg,  which is strongly  absorbed (N$_\mathrm{H}\sim 3\times10^{24}$ cm$^{-2}$) and
has an intrinsic luminosity  in the QSO regime  (L(2--10 keV) $\sim 3 \times 10^{44}$\lum).\\ The
2-- 10 keV spectrum of \sorg\ is remarkably hard ($\Gamma\sim 1.2$), and
presents a strong iron line (EW $\sim$ 0.7 keV), clearly suggesting that below 10
keV we are seeing only reprocessed radiation.  Since the energy of the Fe K
emission is found to be at $\sim 6.7$ keV, consistent with He-like Fe, and its EW
is too high to be explained in a starburst dominated scenario, we suggest that the
2--10 keV emission of IRAS 19254-7245 is  dominated by reflection/scattering from
highly ionized matter. Indeed, within this latter scenario we found that the
photon index of the illuminating source is $\Gamma=1.87^{+0.11}_{-0.28}$, in
excellent agreement with the mean value  found for  radio quiet unobscured AGN. }
{}
{}
{}{}

 \titlerunning{A Suzaku observation of the ULIRG IRAS19254-7245}              

   \maketitle
%

\section{Introduction}

Ultra Luminous Infrared Galaxies (hereafter ULIRGs)   are  an enigmatic class of
sources  which emit most of their energy  in the far-infrared (FIR, 8--1000$\mu
m$) domain \citep{Sanders},  with  luminosities above $\sim 10^{12}$ L$_\odot$,
(i.e., comparable to  QSO luminosities).   The importance of  understanding the 
physical processes at work in ULIRGs  is strengthened by the observational
evidence  that they are generally advanced mergers of gas-rich galaxies; these
events are now   considered to be at the origin of some of the massive elliptical
and S0 galaxies \citep{hopkins2005, hopkins2006, springel05} and the QSO stage
could be a phase during the  evolution of these systems.  However,  understanding
their physical  nature is complicated by the large amount of obscuration from
dust present in these sources, which makes it difficult to directly observe the
nuclear source.\\

  X-ray observations    of ULIRGs performed  with XMM-Newton
\citep{France, Braito} and Chandra \citep{Ptak,Teng05} and recently Suzaku 
\citep{Teng} have proved to be a
fundamental tool to disentangle the contribution of starburst and AGN activity
and to investigate the presence  of hidden AGNs in these sources.   These 
observations have shown that ULIRGs   are intrinsically faint X-ray sources, 
their observed X-ray luminosities being typically L(2--10 keV)\ltsima
$10^{42}-10^{43}$\lum.    The   X-ray spectra  of  ULIRGs  are
complex  and present the signatures of both the starburst  and the AGN activity,
confirming the composite nature of ULIRGs.   
These studies have
also shown  that   more than half of the     local brightest  AGN-ULIRGs  (5/8) host an obscured AGN, with three 
being Comption Thick (N$_H >
10^{24}$cm$^{-2}$; NGC6240, \citealt{6240}; Mrk~231, \citealt{231} and UGC5101,
\citealt{5101}).  Observations  above 10
keV  are thus fundamental for measuring the intrinsic X-ray luminosity  of obscured
AGN hosted in ULIRGs  and its  contribution to their high observed FIR
emission.\\

 IRAS 19254-7245 (also known as the {\it SuperAntennae}) belongs to a flux
limited   sample at  $60\mu$m composed of  the 15  brightest nearby ULIRGs
\citep{Genzel}. Located at $z=0.062$, \sorg\ has an infrared luminosity of
$L_\mathrm{8-1000\mum}=1.1\times 10^{12}L_\odot$ corresponding to a bolometric
luminosity of $L_{bol}\sim 4\times 10^{45}$erg s$^{-1}$. Like most of the ULIRGs,   IRAS 19254-7245 is a
merger system of two gas-rich spiral galaxies. 

 The southern nucleus,  optically classified as a Seyfert 2, is one of the
brightest nearby ULIRGs which has proved to host both a powerful starburst and an
obscured AGN, while there is no evidence of AGN activity in the northern
nucleus.\\ A previous X-ray observation  of \sorg\  performed with  \xmm\ 
suggested that this ULIRG   harbors a heavily obscured   and high-luminosity
AGN. Indeed the hard power-law continuum above 2 keV   (photon index $\Gamma
=1.3$) and the detection of a strong Fe-K$\alpha$ emission line at $6.5\pm 0.1$
keV ($EW \sim 1.4$ keV) were highly  indicative of a Compton-thick source
\citep{Braito}.   As the   two nuclei  are located $\sim$ 9 arcsec  apart from each
other, \xmm\ could not resolve them; however the centroid of the hard X-ray emission 
was   spatially coincident with the southern  nucleus.   Chandra observation  which would settle or solve this
issue  has not  been performed yet. \\  
The  best fit model obtained
for the 0.5-10 keV  X-ray emission detected with \xmm\ was composed by a strong
soft thermal component, associated with the starburst emission and a hard X-ray
 component associated with the AGN activity. This AGN component was parametrized with
a  Compton thick  AGN model  and was composed of a  pure Compton-reflected continuum (with $\Gamma\sim
1.8$), a scattered power law component and a strong Fe emission line.  The observed 2--10
keV luminosity  of the AGN was found to be   $\sim 4\times 10^{42}$ \lum.  Due to the limited
 energy bandpass of XMM-Newton, this observation did not allow  us to  directly see the intrinsic continuum, thus to   measure
the  absorbing column density and  the intrinsic X-ray luminosity of
IRAS~19254-7245. \\

Here we present
the analysis of a deep Suzaku observation  ($\sim 150$ ksec) of this system,
which allowed us  for the first time to   constrain the intrinsic  power of the
AGN hosted in \sorg, as well as to investigate in detail the properties of the
Fe line complex. In Sec.~2  we present the Suzaku data analysis and results,
while in Sec.~3 we  discuss the overall scenario for X-ray emission of \sorg.
Throughout this paper,  the current popular
cosmology is assumed with $H_0=73$ km s$^{-1}$ Mpc$^{-1}$, $\Omega_{\rm{M}}=0.27$
and $\Omega_\lambda=0.73$.

\section{Observations and data reduction}

Suzaku  (\citealp{Mitsuda}) is the fifth Japanese  X-ray satellite, which  carries on board  four sets
of X-ray mirrors, with a X-ray CCD  (XIS; three front illuminated, FI,  and one
back illuminated, BI \citealp{XISref}) at their focal plane, and a non imaging hard X-ray
detector   (HXD, \citealp{Takahashi}). The latter is composed by 2 main instruments: the Si PIN
photodiodes and the GSO scintillator counter. Altogether the XIS and the
HXD-PIN cover the 0.5--10 keV and 12--70 keV bands respectively.\\

Suzaku     observed IRAS19254-7245 for a total exposure time of about 150~ksec;  the
observation was performed at the beginning of November 2005, when all the 4 XIS  were
still working\footnote{Later that month Suzaku XIS2 failed. No charge injection (see:
http://suzaku.gsfc.nasa.gov/docs/suzaku/analysis/sci.html) was applied at the time of the
observation  so the nominal energy resolution of the XIS at 6 keV was degraded with
respect to the prelaunch one}.\\ 
Cleaned event files from the version
2 of the Suzaku pipeline processing were used with the  standard screening\footnote{The
screening filter all  events  within the South Atlantic Anomaly (SAA)  as well as  with an
Earth elevation angle (ELV) $ < 5\ensuremath {{}^{\circ }}$ and  Earth day-time elevation
angles (DYE\_ELV) less than $ 20\ensuremath {{}^{\circ }}$. Furthermore also data within 
256 s of the SAA were excluded from the XIS and within 500s of the SAA for the HXD. Cut-off
rigidity (COR) criteria of $ > 8 \,\mathrm{GV}$ for the HXD data and $ > 6 \,\mathrm{GV}$
for the XIS were used.}. The net exposure times are   $97.9$ ksec  for each of the XIS
and  $142.1$ ksec for the HXD-PIN. The XIS  source spectra  were extracted from a circular
region of 2.9$'$ radius  ( which correspond to an energy encircled fraction of
90\%\footnote{see ftp://legacy.gsfc.nasa.gov/suzaku/doc/xrt/suzakumemo-2008-04.pdf})
centered on the source. Background spectra  were extracted from two circular regions of
2.4$'$ radius  offset from the source and the calibration sources.  The XIS response
(rmfs) and ancillary response (arfs) files were produced,   using the latest calibration
files available, with the \textit{ftools} tasks \textit{xisrmfgen} and
\textit{xissimarfgen} respectively.  The net 0.5--10 keV  count rates  are: $(1.67\pm
0.07)\times 10^{-2}$ cts/s, $(1.57\pm 0.06)\times 10^{-2}$ cts/s, $(1.55\pm 0.06)\times
10^{-2}$ cts/s and $(1.84\pm 0.09)\times 10^{-2}$ cts/s for the  XIS0, XIS2, XIS3 and
XIS1  respectively.  The source spectra from the three FI CCDs were then combined, while
the  BI (the XIS1) spectrum  was kept separate and fitted simultaneously.  The net XIS
source spectra were then  binned  in order to have a minimum S/N of 4  in each energy bin
and   $\chi^2$ statistics have been used. 

\subsection{HXD-PIN data reduction} For the HXD-PIN data  reduction and analysis
we followed the latest Suzaku data reduction guide (the ABC guide Version
2)\footnote{http://heasarc.gsfc.nasa.gov/docs/suzaku/analysis/abc/}.  For the
analysis we used the rev2 data, which include all 4 cluster units, and the 
best background available \citep{fukazawa}, which account for  the instrumental background
(NXB; \citealp{Takahashi,kokubun}). We  then
simulated a spectrum for the cosmic X-ray background counts \citep{Boldt,Gruber} and
added  it to the  instrumental one.\\ 

At the time of the writing two different  instrumental background files have
been released  (background A or ``quick''  background and the background D or
``tuned'' background; Mizuno et al.
2008\footnote{http://www.astro.isas.jaxa.jp/suzaku/doc/suzakumemo/suzakumemo-2008-03.pdf};\citealt{fukazawa}).
We  tested both the instrumental backgrounds and we included a $\pm 10\%$ uncertainty
in the level of the CXB. The inspection of the  \sorg\  net spectrum shows that the
source is detected  in the 15--30 keV  with both the two background files. The net
count rate in the 15--30 keV using background A and D are respectively 
$1.59\pm0.14\times10^{-2}$ cts s$^{-1}$ and  $1.48\pm0.14\times10^{-2}$ cts 
s$^{-1}$ and the corresponding background count rates   are $0.25\pm 0.004$ cts s$^{-1}$ 
and $0.26\pm 0.003$ cts s$^{-1}$.\\
We then decided to use the latest
release (background D), which  is affected by lower
systematic uncertainties (of about 1.3\% at 1$\sigma$), which correspond to  about half of the  first
release of the NXB.  
Using this background \sorg\ is detected in the 15--30 keV band  at $\sim 5.5$\%  above the
background (a total of  $\sim$ 2000 net counts have been collected),
corresponding to a signal-to
noise ratio $S/N\simeq 10.8$.  The dominant component in the background is the instrumental one with a
count rate of $0.24\pm 0.001$cts s$^{-1}$, while the CXB count rate ranges from $1.4 \times 10^{-2}$
to  $1.6 \times 10^{-2}$ when we  include the $\pm 10\%$ uncertainty on its level.
If we  then assume a 10\% higher CXB level the source is still detected  at
$5.0$\% above the background (mean count rate  in the 15--30 keV is 
 $1.35\pm0.14\times10^{-2}$ cts s$^{-1}$) with S/N$\simeq 9.7$, thus the detection of
 \sorg\ is not dependent on the CXB absolute level.
 As a further check for the level of the CXB we analyzed the  Suzaku observation
of the Lockman Hole performed   in May 2007. We performed an identical analysis of the CXB
HXD-PIN observation as we did for \sorg. 
The flux  of the CXB measured with the Lockman Hole  observation is  F(15--50 keV)$=
1.1\pm0.2\times 10^{-11}$ \flux\ (corresponding to a  flux density of $3\times 10^{-11}$\flux
deg$^{-2}$), in  agreement with the flux  of the simulated CXB (F(15--50 keV)$=1.0\pm0.2\times
10^{-11}$ \flux) and with the flux measured with
BeppoSAX (\citealt{Frontera})  and the recent measurement obtained with {\it Swift}
(\citealt{Moretti}). \\

  Since the HXD-PIN is a non
imaging detector, and  taking into account the large field of view 
 of the instrument (0.56 deg $\times$ 0.56 deg), we first checked that the 
detection is not due to another X-ray source.  In particular,
we searched  the NED data base  for known AGN  in the HXD field of view
and we inspected the available \xmm\  observation.   Indeed, two  X-ray
sources with a 2--10 keV flux comparable to \sorg's emission  are 
detected with \xmm.  The two sources are both AGNs, belonging to 
the XMM Bright Survey sample (XBS~J193138.9-725115 and XBS J193248.8-723355;
\citealt{Della Ceca04, Caccia08}). XBS~J193138.9-725115 is a type 1 AGN
(z=0.701) and its  \xmm\ spectrum is well modeled with  single unabsorbed power law 
component ($\Gamma\sim 2$) with no evidence of absorption. XBS J193248.8-723355 is
a Seyfert 2 at z=0.287; the X-ray emission of this source  is in
agreement with the classification as a Compton-Thin Seyfert; indeed,  a
low energy cut-off is present  in the \xmm\ spectra corresponding to 
$N_H\sim 10^{22}$cm$^{-2}$ and there there  is no evidence that  this
source  could be Compton thick.
The predicted 15--30 keV emission   from these sources (derived from the analysis of the \xmm\ and
Suzaku data)  is  less
than $\sim 10^{-13}$ erg cm$^{-2}$ s$^{-1}$, which is  below the HXD-PIN
sensitivity and a  factor of $\sim 50$  below  the measured 15--30 keV flux. \\

For the spectral  analysis we  rebinned the HXD-PIN spectrum of \sorg\   to have a
signal-to-noise ratio of  5 in each energy bin.  In order to have a first  estimate of the
15--30 keV flux  and luminosity of \sorg\, we fitted the HXD-PIN spectrum assuming  a power
law  model  with $\Gamma=1.9$  (i.e. a standard AGN value;  
\citealt{Reeves00,Page04,caccia04}).  Taking
into account the systematic uncertainties of the NXB model,  with this simple model we 
obtained    F(15--30 keV)$\sim 5.2\pm 1.1\times 10^{-12}$ \flux,  and L (15--30 keV)$\sim
4.7\times 10^{43}$ erg s$^{-1}$.  The extrapolation of this model down to 2 keV predicts an
intrinsic luminosity (which does not include a correction for  Compton scattering) of the AGN
in \sorg\ of L(2--10 keV)$\sim 9.5\times 10^{43}$ erg s $^{-1}$. \\

\subsection{The broad band spectrum}

Overall the Suzaku observation confirms the \xmm\ results. A good fit for the 0.5--10
keV Suzaku data is obtained with a  model composed by:  a thermal emission component
($kT=0.64\pm 0.10$ keV and abundance $Z=Z_\odot$, likely associated with the starburst activity), a strong hard
power law component   ($\Gamma=1.2\pm 0.1$, likely associated with the AGN emission),
and a strong  iron  K emission line  ($EW= 710^{+190}_{-170} $ eV, with respect to the
observed continuum). The flux and  observed luminosity, [F(2--10 keV) $\sim 2.9\times
10^{-13}$ \flux, and L(2--10 keV) $\sim 2.5\times 10^{42}$\lum], are found to be
consistent with the values measured     with \xmm. The de-absorbed luminosity of the
starburst component is L(0.5--2 keV) $\sim 4\times 10^{41}$\lum\ also in agreement with
luminosity measured with \xmm. \\

 We  then  tested the best fit model obtained for the \xmm\  spectrum. In this model the  
soft X-ray emission is still modeled with a thermal component,  while the    hard ($\Gamma\sim
1.2$) power law component is replaced with a pure Compton reflected continuum (the  {\sc
pexrav} model in Xspec, \citealp{pexrav}, with an intrinsic  $\Gamma=1.8$) combined with a
moderately absorbed ($N_{\rm H}\sim 5\times 10^{21}$cm$^{-2}$) power law component with the
same $\Gamma$, representing the possible scattered emission. The parameters of the  reflection
component are: an inclination angle $i=45^\circ$, abundance $Z=Z_\odot$ (using the abundances
of \citealt{wilms}) and  a   reflection fraction  (defined by the subtending solid angle of the
reflector $R=\Omega/2\pi$)  R fixed to 1. The normalization of this component  was allowed to
vary. This model is a good fit to the XIS data alone ($\chi^2/\mathrm{dof}=159/144$, see Fig.
1) and the fluxes and observed luminosity of the AGN components are consistent with the values
measured with the \xmm\ observation.   However, this model clearly  under predicts the counts detected
above 10 keV (see Fig. 1 green data points). Indeed,  when we include  in the fit the HXD-PIN  
data, fixing the  cross-normalization between the XIS  and the PIN  to 1.16 (Manabu et al.
2007; Maeda et al.
2008\footnote{http://www.astro.isas.jaxa.jp/suzaku/doc/suzakumemo/suzakumemo-2007-11.pdf;\\
http://www.astro.isas.jaxa.jp/suzaku/doc/suzakumemo/suzakumemo-2008-06.pdf}), the model is
statistically  unacceptable ($\chi^2/\mathrm{dof}=231/147$) and  even allowing for  a harder
photon index  ($\Gamma\sim 1.3$) it  is not a good representation of  the   0.5--30 keV 
emission  ($\chi^2/\mathrm{dof}=220/146$).\\ 

\begin{figure}
\begin{center}
\includegraphics[angle=-90,width=0.5\textwidth]{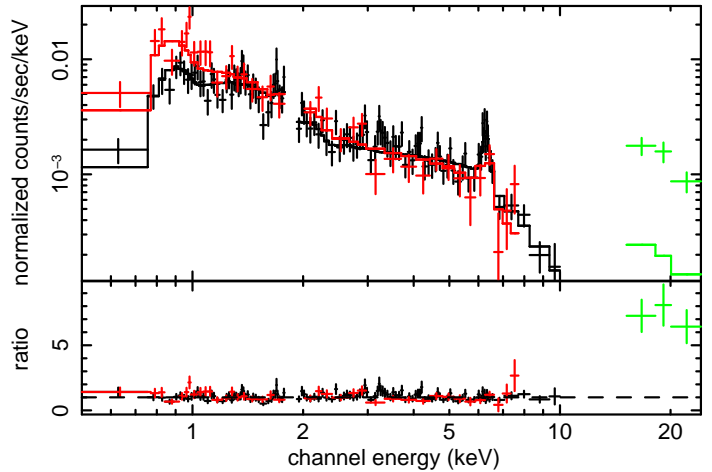}

\caption[paper_newfig1.eps]{ 0.5-30 keV Suzaku XIS-FI (black data points), XIS-BI (red) and
HXD-PIN
(green) data and ratio with the 0.5-10 keV  best fit model. The model is composed by: a thermal
emission component dominating the 0.5-2 keV emission, a Compton reflected continuum (R=1,
$\Gamma=1.8$) and a strong Fe emission line at 6.7 keV (EW$\sim 700$ eV). A clear excess is present above 10
keV, which we attribute to the intrinsic X-ray emission of \sorg\ transmitted through the high column
density absorber. 
  \label{fig:xmmmo}}
\end{center}
\end{figure}

In order to  account for the  excess detected  above 10 keV, we added to the model  a
second heavily absorbed power law component. Since the HXD/PIN residuals suggest the presence of
a high  column density absorber,  we used for   this component the  model by
Yaqoob (1997) ({\sc plcabs} in Xspec),  which correctly takes into account Compton 
down-scattering. Indeed, for high column densities  the observed  X-ray continuum  may also be suppressed  by
Compton down-scattering,  and the intrinsic luminosity  must be corrected by a factor $e^{\tau}$
, where  $\tau=N_{\rm H}\sigma_\tau$ and  $\sigma_\tau = 6.65 \times 10^{-25}$cm$^{2}$ is the
Thomson cross-section. \\
 
This model  provides now a good fit for the 0.5--30 keV spectrum
($\chi^2/\rm{dof}=164/145$; see Fig.~2 upper panel and Table 1 model A). However,  we found a
low value of the reflection fraction with respect to this primary absorbed power law component
($R<0.1$). This suggests  that the line and the hard   2--10 keV
spectrum  are   unlikely to be   produced by reflection off cold material. Indeed, the   broad band 
continuum   could be also reproduced  by  a model   without the reflected component (see Table 1 model B).
Statistically this model  gives a  slightly  worse fit ($\chi^2/\rm{dof}=181/146$) than the previous one, 
but it is not able to account for the hardness of the  2--10 keV emission. In particular  clear residuals 
are  present in the 5--10 keV band  where the reflected component  dominates (see Fig. 2 top panel). 
Finally,  if we allow the photon index to vary,   we found that although the fit
improves  we again  need an unusually  hard  photon index of the power law component
($\Gamma=1.20^{+0.11}_{-0.04}$, $N_{\rm H}=4\pm 1\times 10^{24}$ cm$^{-2}$; see Table 1 model C).\\

\begin{figure}
\begin{center}
\includegraphics[angle=-90,width=0.5\textwidth]{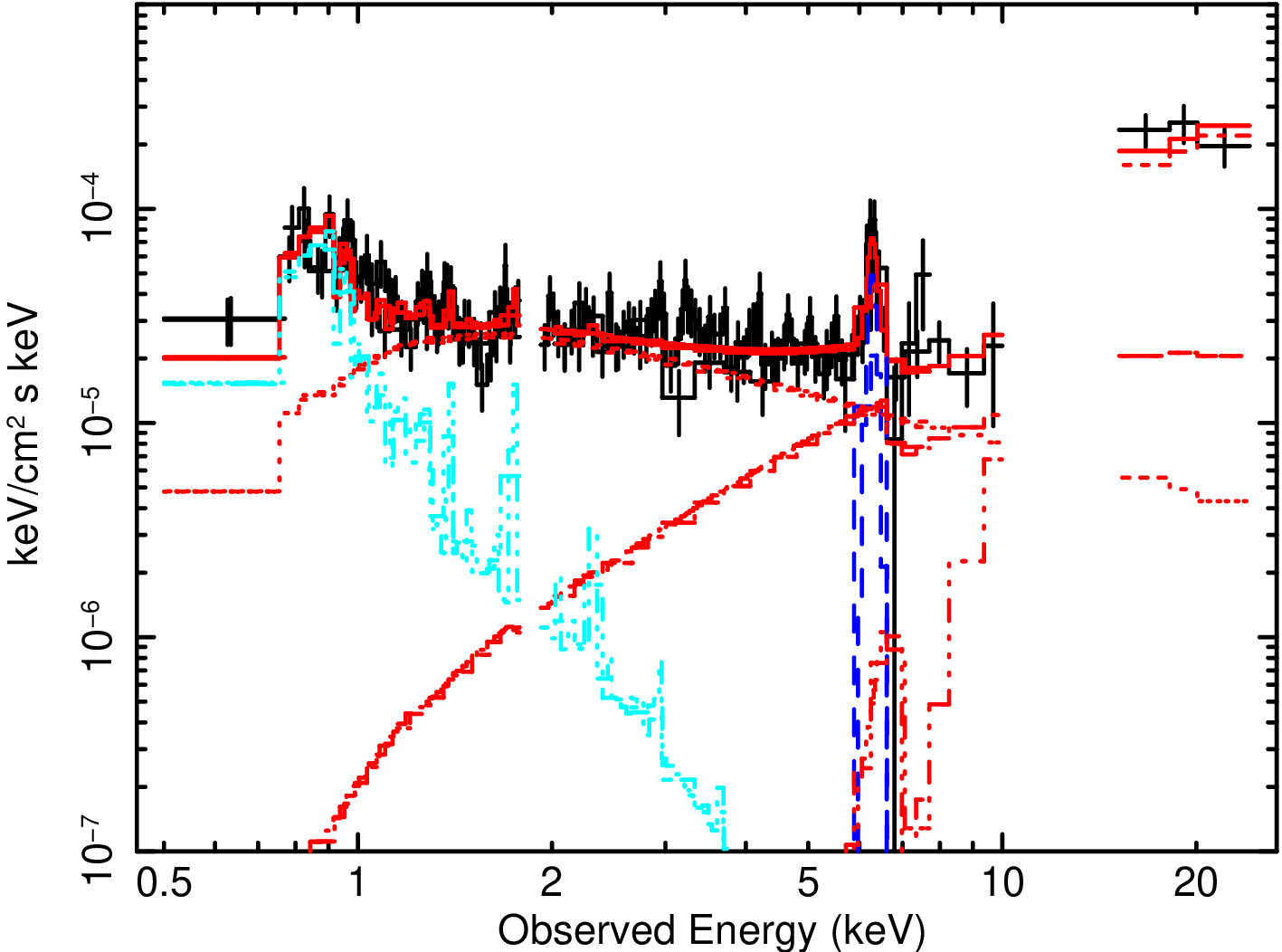}
\includegraphics[angle=-90,width=0.5\textwidth]{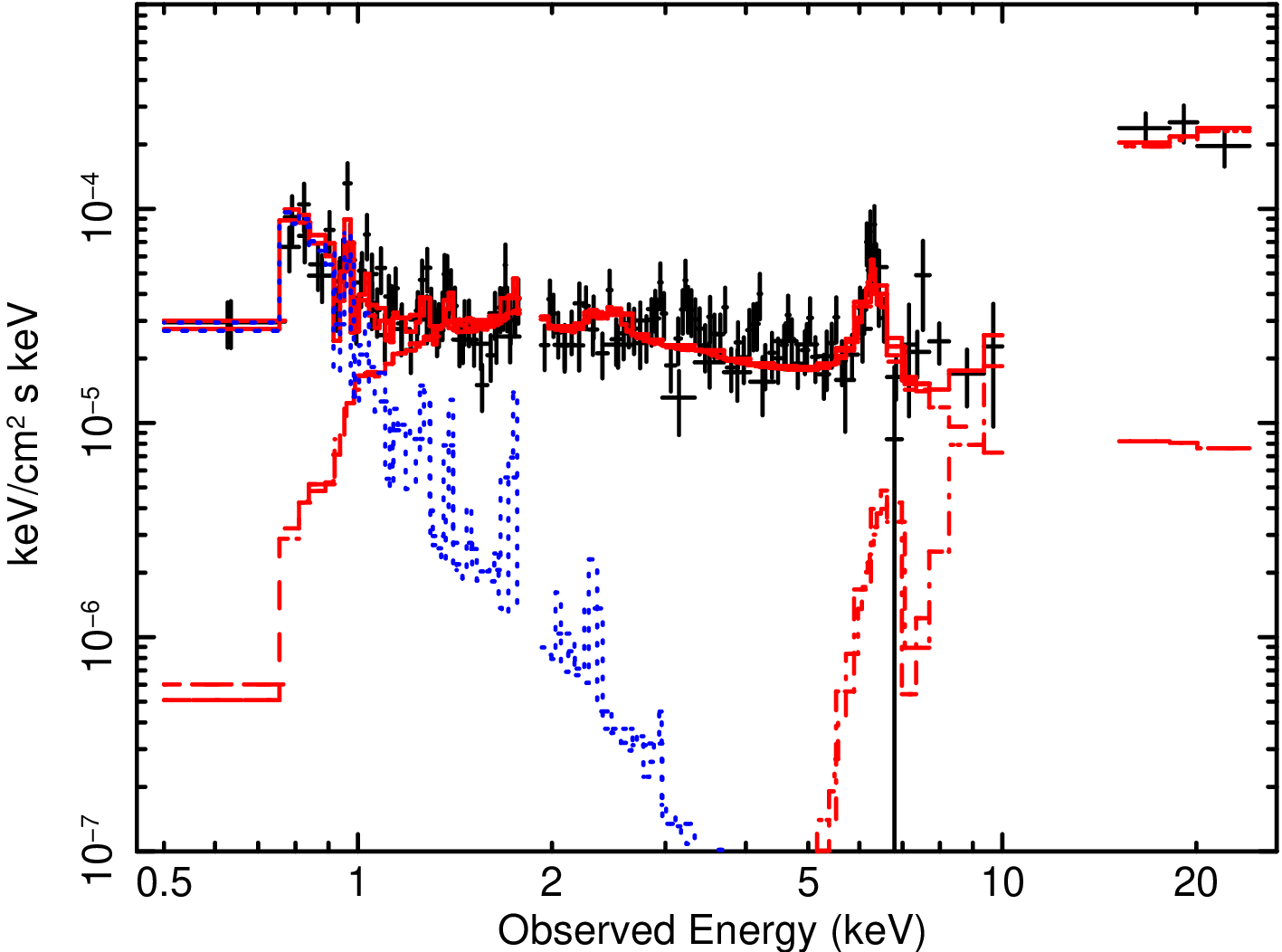}

\caption[new_fig2.eps]{Upper panel: Suzaku  XIS and HXD spectra of \sorg\ when the underlying AGN continuum is 
modelled with a neutral Compton-reflected component and  power-law component seen in transmission through a
high column density absorber (N$_{\rm H}\sim 4\times10^{24}$ cm$^{-2}$). Lower panel: same as above but with
the 2--10 keV AGN emission  modelled with an ionized reflected  component (the high energy component is
modelled as in the upper panel with a power-law component seen in transmission through the high column density
absorber).
 \label{fig:fig2.ps}}
\end{center}
\end{figure}

 One of the main  results of this observation is that, although we can confirm the presence of a strong
Fe line as detected with  \xmm, the centroid of this line is  now at $ 6.67\pm 0.05 $ keV in
the rest frame   ($EW\sim 0.7$ keV; for the model without the reflected component), consistent with
He-like Fe.   Furthermore,  the line appears to be  marginally broad
($\sigma=0.12\pm 0.06$ keV).  The inclusion of the line in the model improves the fit by 
$\Delta\chi^2= 51$  for 3 degree of freedom\footnote{for the scenario with the  reflected
component  the fit improves by  $\Delta\chi^2= 40$}.  However, if we constrain the line to be
unresolved  the fit is worse only by
$\Delta\chi^2=5$. In order to check  the energy and the intrinsic  width of the Fe line detected
in the XIS, we examined the spectra of  the $^{55}$Fe calibration source lines, which are located
on  two corners of  each XIS. The calibration source produces lines from Mn $
\mathrm{K}\alpha_{1}$  at 5.899 keV and  Mn $ \mathrm{K}\alpha_{2}$ at 5.899keV.  From the spectrum
of the calibration source  we  found that the line energy is shifted red-wards  by  about 25 eV,
while the residual width is $\sigma \sim 50 $eV\footnote{This residual width is due to the degradation of the
XIS after the launch and prior to the correction with the   charge injection}; which confirms that  the broadening of the line is intrinsic to
the source and not  instrumental\footnote{ The intrinsic width of the Fe line in the spectrum of
\sorg\ can be   $ \sigma_{\mathrm{int}}^{2} = \sigma_{\mathrm{meas}}^{2} -
\sigma_{\mathrm{lamp}}^{2}$ (where $ \sigma_{\mathrm{meas}}$ is the measured width and $
\sigma_{\mathrm{lamp}}$ is the width of the calibration lines)}. After the subtraction  in quadrature of this residual
width we get  $ \sigma_{\mathrm{int}}^{2} =110\pm60 $ eV).  However, taking into account  the
present count statistics of the  data, this broadening could be  due to the presence of other line
components, which are not resolved. In particular,  the line profile can be    explained  with
three unresolved  Gaussian lines (at  6.4 keV, 6.7 keV and 6.96 keV; see Fig. 3).    Statistically
this model gives a similar good fit than  models with a single broad line
($\chi^2/\rm{dof}=161/145$)  with the strongest line  being  the 6.7 keV line ($EW\sim 400$
eV). Though  the other two lines are not  statistically required  a weak  ($EW< 200$ eV) emission
line could be present at the energy of the neutral Fe K$\alpha$, while the 90\% upper limit on the
6.96 keV line is $150$ eV.  \\

\begin{figure}
\begin{center}
\includegraphics[angle=-90,width=0.5\textwidth]{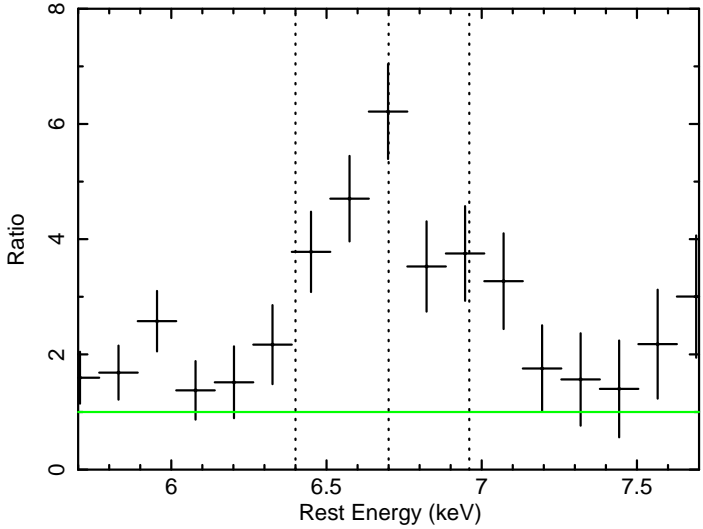}

\caption[paper_newfig4_rest.eps]{Residuals  of the data/model of \sorg\ XIS  data at the Fe band,
when no iron line is included in the model. The three vertical lines  highlights the   energy
centroids  of  the  three possible components of the Fe line complex (6.4 keV, 6.7 keV and 6.97 keV 
). The energy scale is in the rest frame.\label{fig:fig3.ps}}
\end{center}
\end{figure}

Since the energy centroid of the line detected with Suzaku appears to be in disagreement
with the \xmm\ results and the EW appears to be lower we  went back to the \xmm\ data and compared 
them  with the Suzaku results. The exposure time of the \xmm\ observation was only 20 ksec and when we
take  into account  the errors on the  flux and line continuum   we found that the  two lines are
consistent within each others.  Furthermore,    the energy centroids are consistent within the errors
($E_{XMM}=6.5\pm0.1$ keV; $E_{\mathrm {Suzaku}}=6.66\pm 0.05$ keV).  Finally, a possible blending of 3
lines was also present in the \xmm\ data, but again the low exposure time of this
observation does not allow a more detailed analysis of the Fe line profile.

From  a statistical point of view    all these  models are a good representation of the 0.5--30
keV emission, but they  are unable to account for the  hardness of the continuum.  Furthermore
the  line energy  of the strongest emission line is at odds with a scenario where the 2--10 keV
emission is dominated by reflection/scattering  off cold material as assumed with the  continuum
model tested above.   One  possibility is that the 6.7 keV line is due to reflection from  highly ionized matter;  we
thus replaced the cold reflected ({\sc pexrav}) power law component with an ionized reflected
component, as  is described  by the the \citet{Ross} table (otherwise known as the {\sc reflion}
model). This model allows different values for the ionization    parameter of the reflecting
material and   it also includes the Fe  K emission line, as well as  emission lines from other
elements in addition to the reflected continuum.  We fixed  the Fe abundance to solar and we
included a lower column density in front of the reflected component. The photon  index of the
illuminating X-ray source is left as a free parameter.  A good fit  (see Fig.~2 lower panel and
Table 1 model D) of the 0.5-30 keV emission is  obtained  with a ionization parameter of $\xi=
1000^{+170}_{-430}$ erg cm s$^{-1}$, where the value of the ionization is determined mainly by the
strength and energy of the Fe line; at this ionization level, Fe K emission is    almost entirely
due  
to Fe xxv.  The  reflected component is modified by a lower column density absorber with   $
N_{\mathrm{H}}\sim 10^{22}$ cm$^{-2}$, which  is  probably  on a  larger scale
 than the inner  high column density absorber.
 We stress that a second   highly  absorbed ($N_{\rm{H}}\sim
3\times10^{24}$cm$^{-2}$) power-law component is still required to account for the
HXD-PIN emission  and the intrinsic 2--10 keV luminosity  is $\sim 3\times
10^{44}$\lum.  The photon  index of the illuminating source is now   
$\Gamma=1.87^{+0.11}_{-0.28}$, consistent with the mean  value measured  in
unobscured radio quiet AGN ($\Gamma_\mathrm{mean}=1.9$; \citealt{Reeves00}).  \\

\begin{table*}
      \caption[]{Results of the Spectral Fit}
      \label{tab:fits}
      \begin{centering}
     \renewcommand{\footnoterule}{}  
\begin{tabular}{ccccccccc}
      
 \hline\hline
Model& $\Gamma$ & $N_{\rm H}$  & E$_{\rm c}/\xi$ & $\sigma^{a}$ & EW$^b$ & L(2--10 keV)$^c$&
L(10--30 keV)$^c $&
$\chi^2/dof$\\ 
 &   & $10^{24}$ cm$^{-2}$ & keV/erg cm s$^{-1}$ & keV & keV & 10$^{44}$\lum &
10$^{44}$\lum & \\
\hline
A &  1.8$^{fixed}$ & 4.2$^{+2.7}_{-0.9}$ &6.67$^{+0.05}_{-0.05}$ &  0.11$^{+0.06}_{-0.06}$ & 0.60$^{+0.12}_{-0.22}$ &  4.0   &3.6& 164/145 \\
 B$^{d}$ &  1.8$^{fixed}$ & 3.1$^{+1.2}_{-0.4}$ &6.66$^{+0.04}_{-0.05}$ &  0.14$^{+0.07}_{-0.06}$ & 0.86$^{+0.24}_{-0.25}$ &  2.1   & 1.9&181/146 \\
 C$^{d}$    &1.20$^{+0.11}_{-0.04}$   & 4.1$^{+1.3}_{-1.3}$ &6.67$^{+0.05}_{-0.04}$   &  0.12$^{+0.06}_{-0
.05}$  &0.67$^{+0.16}_{-0.15}$ &   1.7  & 3.9& 154/145\\

D$^{d}$   &1.87$^{+0.11}_{-0.28}$   & 3.2$^{+1.3}_{-0.5}$ &1000$^{+170}_{-430}$   &  $-$ 
 &$-$ &   2.6  & 2.2& 199/147\\
    \hline
  \end{tabular}
  \end{centering}
  
   $^a${ The  values of $\sigma$  are the measured  ones, which are not corrected for 
   width of the calibration lines,  $\sigma_\mathrm{lamp}$} \\
   $^b${ The EW is   measured against the total observed continuum}\\ 
   $^c${ The luminosities are  derived from the XIS front illuminated CCDs}\\
   $^d${  These models do not include a cold reflected component}\\

\end{table*}

In
summary, this model is now able to reproduce  in a consistent way all the main
characteristic of the  the broad band X-ray emission of \sorg\ and  in particular the
Fe emission line detected at $\sim 6.7$ keV and the flatness of  the observed 
continuum. Finally, it is worth noting that, independently of the assumed model for
the 0.5-10 keV emission, we always need to include an  absorbed power law  component
to account for the HXD-PIN data with a $N_{\rm H}\sim 3-4\times 10^{24}$cm$^{-2}$,
and the      derived  2-10 keV  intrinsic luminosity   is always  above L(2--10 keV) 
$\sim 10^{44}$\lum,  ranging from $2\times  10^{44}$\lum  to $4\times  10^{44}$ \lum
(see Table 1). \\

\section{Discussion and Conclusions}

The detection of the  Fe K  emission line at 6.7 keV instead of the  6.4 keV emission  line  expected
from neutral iron may suggest that 
this line is associated with   strong starburst  activity and that the emission below 10 keV is
not due to the AGN, but rather to a hot thermal plasma as expected in a starburst dominated scenario.
Indeed, from a statistical point of view we can obtain a good fit ($\chi^2/dof=147/144$) replacing the
AGN reflected emission with a thermal component. This model gives a best fit temperature of
$kT=8.1^{+1.2}_{-1.3}$ keV, $N_{\rm H}\sim 6\times 10^{21}$cm$^{-2}$, twice solar abundances ($Z\sim
2.1 Z_\odot$) and a luminosity of $L(2-10\; \rm{keV})\sim 2\times 10^{42}$ erg s$^{-1}$. \\

 A possible origin of the high temperature ($kT\sim 8$ keV) plasma could be
the presence of several SNe; their X-ray emission could in principle explain
the high temperature as well as the presence of the strong 6.7 keV line
\citep{persic2002}.  From the FIR  luminosity  we estimated a SFR for \sorg\
of $\sim $200M$_\odot$yr$^{-1}$ and  a SN rate of $\sim 2$SNe yr$^{-1}$
(\citealt{Mannucci}). However, even  assuming that the SNe are at  the higher
end of the expected range of X-ray  luminosity ($L_x=10^{40}-10^{41}$\lum), we  need
a factor 10 times more SNe than the predicted rate to maintain the observed hard
X-ray emission. Furthermore, X-ray observations of SB galaxies  showed that 
the major contributor to the  2--10 keV emission is the integrated emission
from  High-mass X-ray  binaries (HMXB)  and that there is  a linear   relation
between   the SFR and the 2--10 keV   emission from HMXB. In the case of 
\sorg\ if all the FIR luminosity is due the SB activity, then  the predicted
2--10 keV luminosity  of the HMXB is $\sim 10^{42}$ \lum\
(\citealp{Grimm,Ranalli,Persic}).   In this scenario we would thus require
that, contrary to what seen in the other SB, the hot diffuse  emission has a
luminosity similar to  the contribution from  HMXB. In summary, although the
starburst model can well reproduce the line intensity and the overall shape of
the 2--10 keV continuum, the luminosity of this thermal component (L (2--10
keV)$\sim 2 \times 10^{42}$ erg s$^{-1}$) is likely to be too high for a pure starburst
scenario. \\

 Finally, we would like to note that  if we assume that the emission
detected above 2 keV is dominated by the emission of  unresolved HMXB   we
still cannot explain the 6.7 keV emission line.  Indeed,  we would expect a
lower EW of the Fe line (EW$\sim 0.3$ keV; \citealt{white, persic2002}),
which  it is not consistent with the high value observed in \sorg.    One possibility
is   that only a fraction of the Fe line at 6.7 keV  originates in a high
temperature plasma. We thus added to our best fit model  a second high
temperature emission fixing the abundance to  the solar value. This high
temperature thermal emission can account for $\sim 30\%$ of the flux of
the line at 6.7 keV. Though  the inclusion of this emission can account
for   a fraction of the Fe line at 6.7 keV,  we still need  a strong 
ionized reflection component  to account at the same time for the
continuum shape and  the Fe line. \\ 

On the other hand if we attribute the  hard X-ray emission to the presence of the
AGN the  flatness of its observed  continuum   together with strong Fe
emission lines are  usually  considered evidence of a Compton-Thick  source,
where no direct emission is seen below 10 keV and the shape is produced by
reflection off cold matter. However,  contrary to what seen in other Compton
thick sources here  the Fe K emission is predominately in the He-like state   
with little or no 6.4 keV line.\\

It is often assumed that a hallmark of Compton-thick X-ray sources is a prominent Fe~K$\alpha$
emission line at $\sim 6.4$~keV, from neutral matter, with an EW that can exceed a keV.
However, we show below that in practice, the Fe~K$\alpha$ line EW can be significantly
reduced and rendered undetectable for large column densities of the order of $10^{25} \
\rm cm^{-2}$ or more. Now, the observed EW of the Fe~K line is largest when only the
reflected continuum that is associated with the production of the line is observed. 
However, if any additional continuum is observed and if its magnitude is comparable to,
or greater, than the scattered/reflected continuum, the Fe~K line will be ``diluted'' and
the EW reduced.  
This is because for lines of sight that intercept the reprocessor,
the absolute luminosity of the 6.4 keV Fe K line and its associated
reflected continuum can be much less than the continuum from the Compton-thin
scattering zone, so that the EW of the line measured against the total
observed continuum can be much less than that measured against the
continuum emerging from the Compton-thick reflector.

In this scenario the ionized Fe emission line could be  produced in   the same Compton-thin
zone that scatters a fraction of the intrinsic continuum into the line-of-sight. If this
scattered continuum is the dominant observed continuum (as would be the case if the direct
line-of-sight is obscured by Compton-thick matter), then the EW of the emission lines from
ionized Fe could be large, of the order of hundreds of eV or more. The details depend on
several factors, principally the ionization parameter and column density of the
line-emitting region, as well as the shape of the ionizing
continuum (see for example, detailed calculations of the EW of the Fe~{\sc xxv} and Fe~{\sc xxvi}
emission lines in \citealt{Bianchi2002}).    Suppose that the
Compton-thick reprocessor subtends a solid angle $\Delta \Omega/4\pi$
at the X-ray source and that the system is observed along
a line-of-sight that does not give a direct view of
the X-ray source and that has the greatest column density, $N_{H}$,
(this maximizes the Fe~K line EW).
Further suppose that the space between the reprocessor is
filled by a warm, optically-thin scattering zone with Thomson
depth  $\tau_{\rm thin}$ (with $\tau_{\rm thin} <<1 $), subtending a solid angle $1-(\Delta \Omega/4\pi)$
at the X-ray source. Then, a fraction $f\equiv \tau_{\rm thin}[1-(\Delta \Omega/4\pi)]$ 
of the intrinsic X-ray continuum is
scattered into the observer's line-of-sight and will reduce the EW
of the Fe~K line if it dominates over the zeroth-order continuum
that is observed directly through the Compton-thick absorber.
Specifically, for column densities greater than a few $\times
10^{24} \ \rm cm^{-2}$, even a small value of $f$ can
significantly reduce the EW of the Fe~K line. For example, for a
column density of $10^{25} \ \rm cm^{-2}$, a scattering fraction
of $f>0.01$ will reduce the EW of an Fe~K line by more than an
order  of magnitude, so that an EW of 1~keV would be reduced to
less than $\sim 100$~eV, and it could render the line undetectable
(see \citealt{Ghisellini}).  \\
Furthermore,  the  intrinsic EW (i.e. prior to the dilution effect) of Fe K lines depends on several
factors not only the column density of the absorber but also  the geometry of the   absorber
(e.g.  the half opening angle of the  putative torus; see
\citealt{Ghisellini,Ikeda,Matt}). For example for an half opening angle of $30^{\circ}$ and our estimate of the
column density of the neutral absorber the intrinsic EW of the 6.4 keV Fe  line can span the range 
from 1 to 4 keV \citep{Ghisellini}.\\ 
The inferred column
density of the Compton-thick reprocessor 
implies that a scattering fraction of only $\sim 0.1\%$ in the
optically-thin zone is required to  begin to dilute the
6.4~keV Fe~K line and a scattering fraction of a few
percent is sufficient to reduce the EW of the line well
below 100~eV, consistent with the upper limit of the EW  
 and   the $\sim 2$\% scattering fraction    (measured with respect to
the de-absorbed primary power law component) as measured   with the Suzaku data. \\

A second possible geometry  is that we have  a
direct view of the inner surface of the Compton-thick reprocessor, but the outer part of this
reprocessor is ionized. In this case, if the remaining part of the reprocessor is Compton-thin, the EW
of the 6.4 keV Fe K$\alpha$ line will be reduced and  the  emission detected below 10 keV  is the
reflected emission from this inner ionized surface of the torus which will also produce a strong 6.7
keV line. \\ Thus, we
see that the lack of a large EW neutral Fe~K in
IRAS~19254$-$7245 is not unexpected.
The spectrum below 10~keV is then
dominated by this optically-thin scattered continuum and
the dominance of the emission line from ionized Fe is
consistent with this  picture.\\

It is worth noting that  this is not a unique case of a 
detection of a strong 6.7 keV line in a luminous infrared
galaxy. Other examples are   Arp299 \citep{ballo04},
Arp220 \citep{iwasawa05,Teng}, and IRAS 00182-7112
\citep{Nandra07}. For all these sources although the
optical spectra show no  clear signature  of AGN
activity,  their   X-ray emission and the 6.7 keV line can
be explained with  the presence of an AGN  and an ionized
reflector as in the case of \sorg.  However, while in the 
case of  Arp~299 and Arp~220,   the X-ray luminosity is
not indicative  that the  major contributor to the 
bolometric luminosity   is a high-luminosity AGN, in the
case of IRAS 00182-7112  the X-ray luminosity is too large
to be accounted for by  the strong starforming activity
(L$(2-10)> 10^{44}$\lum) as for \sorg.   For all these
sources, the  presence of  a strong ionized Fe line, with
little or no  6.4 keV line,  could be reconciled with the
picture of a heavily obscured AGN assuming that we do not
have a direct view of the reflected continuum,  because 
it  is  is diluted by the scattering from
the ionized matter, that is associated with the production
of the line, or if the surface  of the putative
Compton-thick reprocessor is highly ionized.\\

Overall to account for the  X-ray emission of \sorg\  above
2 keV, we need  two absorbing/reflecting  media: one
neutral and Compton-Thick and  probably  seen in transmission, and 
one ionized and  probably  seen in reflection. This latter is 
responsible for the flat X-ray spectrum emerging below 10
keV, for the He-like Fe K  line and probably  for the  dilution
of the 6.4 keV Fe line produced in the neutral
Compton-Thick absorber.\\

  Despite the various possible models for the 0.5--10 keV emission,
we  always need a neutral high column density absorber to 
account for  the emission emerging above 10 keV.  
Once we  have corrected for the amount of absorption, the
intrinsic  2--10 keV luminosity of the primary AGN
component is  $\sim10^{44}$ erg s$^{-1}$. This high column
density absorber  may also be the one responsible for the
deep hydrocarbon absorption  detected in the  L-band
spectrum \citep{Risaliti2003} at 3.4$\mu m$.   Finally,
independently from the assumed model for the 2--35 keV
emission (i.e.  ionized reflection or scattered power law
component), we always require the soft thermal component
to account for the 0.5--2 keV emission. As already found
for other ULIRGs this thermal component has a temperature
$kT\sim 0.7$ keV; the luminosity of this component is
$L(0.5-2 \;\mathrm{keV})\sim 4\times 10^{41}$ \lum.
Although, we cannot exclude a possible contribution from
the ionized reflector, the measured soft X-ray 
luminosity  is in agreement with that expected  from the
FIR luminosity and the SFR of \sorg\ \citep{Ranalli,Persic}.\\

The total X-ray luminosity estimated from the intrinsic
component at $E>10$~keV can be converted into a bolometric
luminosity and compared with the total infrared emission.
We adopted the $\alpha_{OX}$-luminosity correlation of
\citealt{Steffen} in order to estimate the 2500~\AA\
luminosity, and the \citealt{Elvis}  quasar spectral
energy distribution to estimate the total luminosity of
the AGN component. Assuming
$L_{2-10}\sim10^{44}$~erg~s$^{-1}$, we obtain
$L_{BOL}(AGN)\sim2\times10^{45}$~erg~s$^{-1}$, i.e. about
50\% of the infrared luminosity. A different way to estimate the AGN luminosity is through
its emission in the mid-infrared, the only other band
where the continuum emission of the AGN is not completely
suppressed. \cite{Nardini}, from an analysis of the
Spitzer-IRS spectrum estimated an AGN contribution to the
bolometric luminosity of IRAS~19254-7245 of $\sim25$\%.
Considering the uncertainties in the bolometric
corrections in both the X-ray and the mid-infrared bands,
the two estimates can be considered in rough agreement.
If the difference is assumed to be real, this could be an indication
of the non-complete covering factor of the AGN circumnuclear absorber:
indeed, the estimate form the mid-infrared spectrum is done assuming
a complete reprocessing of the intrinsic AGN emission in the infrared.
However, the optical classification of IRAS~19254-7245 as a Seyfert~2
suggests that the obscuration of the nuclear source is not complete.
The estimates from the X-ray and infrared spectra would be then
perfectly reconciled assuming a covering factor
of the obscuring material of about 50\%.\\

In conclusion,  this deep Suzaku observation    allowed us   to measure for the first time  the
hard X-ray emission of \sorg\ and infer that its  intrinsic 2--10 keV luminosity   is
of about few times $10^{44}$~erg~s$^{-1}$.    We have found  evidence that the AGN hosted in  \sorg\ is
highly obscured, with a measured column density of the
neutral absorber of  $N_{\mathrm H}\sim 3 \times 
10^{24}$cm$^{-2}$.   We confirm the presence of a strong iron K  emission line  with an $EW\sim
0.7$ keV.  The energy of  iron K emission line is found to be consistent with  Fe {\sc xxv}.  
We propose that the   X-ray emission detected below 10 keV can be ascribed to 
scattered/reflected emission from  highly ionized matter, which could be identified with the
warm  Compton-thin gas which fills the  space between the neutral Compton-Thick reprocessor.\\

\begin{acknowledgements}
 VB acknowledge  support from the  UK STFC research council.
RDC acknowledge financial support from the ASI (Agenzia Spaziale Italiana) grant I/088/06/0.
Support for this work was provided by the National Aeronautics and Space Administration through
the NASA grant  NNG04GB78A.
We thank the anonymous referee  for his/her useful comments, which have improved this paper.

\end{acknowledgements}


\begin{thebibliography}{}
\bibitem[Ballo et al.(2004)]{ballo04} Ballo, L., Braito, V., Della Ceca, R., Maraschi, L., Tavecchio, F., \& Dadina, M.\ 2004, \apj, 600, 634 
\bibitem[Bianchi \& Matt(2002)]{Bianchi2002} Bianchi, S., \& Matt, G.\ 2002, \aap, 387, 76 
\bibitem[Boldt(1987)]{Boldt} Boldt, E.\ 1987, \physrep, 146, 215 
\bibitem[Braito et al.(2003)]{Braito} Braito, V., et al.\ 2003, \aap, 398, 107 
\bibitem[Braito et al.(2004)]{231} Braito, V., et al.\ 2004, \aap, 420, 79
\bibitem[Caccianiga et al.(2008)]{Caccia08} Caccianiga, A., et al.\ 2008, \aap, 477, 735 
\bibitem[Caccianiga et al.(2004)]{caccia04} Caccianiga, A., et al.\ 2004, \aap, 416, 901 

\bibitem[Della Ceca et al.(2004)]{Della Ceca04} Della Ceca, R., et al.\ 2004, \aap, 428, 383 
\bibitem[Elvis et al.(1994)]{Elvis} Elvis, M., et al.\ 1994, \apjs, 95, 1 
\bibitem[Franceschini et al.(2003)]{France} Franceschini, A., et al.\ 2003, \mnras, 343, 1181 
\bibitem[Frontera et al.(2007)]{Frontera} Frontera, F., et al.\ 2007, \apj, 666, 86 
\bibitem[Fukazawa et al.(2009)]{fukazawa} Fukazawa, Y., et al.\ 2009, \pasj, 61, 17 
\bibitem[Genzel et al.(1998)]{Genzel} Genzel, R., et al.\ 1998, \apj, 498, 579 
\bibitem[Ghisellini et al.(1994)]{Ghisellini} Ghisellini, G., Haardt, F., \& Matt, G.\ 1994, \mnras, 267, 743 
\bibitem[Grimm et al.(2003)]{Grimm} Grimm, H.-J., Gilfanov, M., \& Sunyaev, R.\ 2003, \mnras, 339, 793
\bibitem[Gruber et al.(1999)]{Gruber} Gruber, D.~E., Matteson, J.~L., Peterson, L.~E., \& Jung, G.~V.\ 1999, \apj, 520, 124 
\bibitem[Kokubun et al. (2007)]{kokubun}Kokubun, M., et al.\ 2007, \pasj, 59, 53 
\bibitem[Hopkins et al.(2005)]{hopkins2005} Hopkins, P.~F., Hernquist, L., Cox, T.~J., Di Matteo, T., Martini, P., Robertson, B., \& Springel, V.\ 2005, \apj, 630, 705 
\bibitem[Hopkins et al.(2006)]{hopkins2006} Hopkins, P.~F., Hernquist, L., Cox, T.~J., Di Matteo, T., Robertson, B., \& Springel, V.\ 2006, \apjs, 163, 1 
\bibitem[Ikeda et al.(2009)]{Ikeda} Ikeda, S., Awaki, H., \& Terashima, Y.\ 2009, \apj, 692, 608 
\bibitem[Imanishi et al.(2003)]{5101} Imanishi, M., Terashima, Y., Anabuki, N., \& Nakagawa, T.\ 2003, \apjl, 596, L167 
\bibitem[Iwasawa et al.(2005)]{iwasawa05} Iwasawa, K., Sanders, D.~B., Evans, A.~S., Trentham, N., Miniutti, G., \& Spoon, H.~W.~W.\ 2005, \mnras, 357, 565 
\bibitem[Koyama et al.(2007)]{XISref} Koyama, K., et al.\ 2007, \pasj, 59, 23 
\bibitem[Magdziarz \& Zdziarski(1995)]{pexrav} Magdziarz, P., \& Zdziarski, A.~A.\ 1995, \mnras, 273, 837 
\bibitem[Mannucci et al.(2003)]{Mannucci} Mannucci, F., et al.\ 2003, \aap, 401, 519 
\bibitem[Matt et al.(1996)]{Matt} Matt, G., Brandt, W.~N., \& Fabian, A.~C.\ 1996, \mnras, 280, 823 
\bibitem[Mitsuda et al.(2007)]{Mitsuda} Mitsuda, K., et al.\ 2007, \pasj, 59, 1 
\bibitem[Moretti et al.(2009)]{Moretti} Moretti, A., et al.\ 2009, \aap, 493, 501
\bibitem[Nandra \& Iwasawa(2007)]{Nandra07} Nandra, K., \& Iwasawa, K.\ 2007, \mnras, 382, L1 
\bibitem[Nardini et al.(2008)]{Nardini} Nardini, E., Risaliti, G., Salvati, M., Sani, E., Imanishi, M., Marconi, A., \& Maiolino, R.\ 2008, \mnras, 385, L130 
\bibitem[Page et al.(2004)]{Page04} Page, K.~L., Reeves, J.~N., O'Brien, P.~T., Turner, M.~J.~L., \& Worrall, D.~M.\ 2004, \mnras, 353, 133 
\bibitem[Persic \& Rephaeli(2002)]{persic2002} Persic, M., \& Rephaeli, Y.\ 2002, \aap, 382, 843 
\bibitem[Persic et al.(2004)]{Persic} Persic, M., Rephaeli, Y., Braito, V., Cappi, M., Della Ceca, R., Franceschini, A., \& Gruber, D.~E.\ 2004, \aap, 419, 849 
\bibitem[Ptak et al.(2003)]{Ptak} Ptak, A., Heckman, T., Levenson, N.~A., Weaver, K., \& Strickland, D.\ 2003, \apj, 592, 782 
\bibitem[Ranalli et al.(2003)]{Ranalli} Ranalli, P., Comastri, A., \& Setti, G.\ 2003, \aap, 399, 39 
\bibitem[Reeves \& Turner(2000)]{Reeves00} Reeves, J.~N., \& Turner, M.~J.~L.\ 2000, \mnras, 316, 234 
\bibitem[Risaliti et al.(2003)]{Risaliti2003} Risaliti, G., et al.\ 2003, \apjl, 595, L17 
\bibitem[Ross \& Fabian(2005)]{Ross} Ross, R.~R., \& Fabian, A.~C.\ 2005, \mnras, 358, 211 
\bibitem[Sanders \& Mirabel(1996)]{Sanders} Sanders, D.~B., \& Mirabel, I.~F.\ 1996, \araa, 34, 749 
\bibitem[Springel et al.(2005)]{springel05} Springel, V., Di Matteo, T., \& Hernquist, L.\ 2005, \mnras, 361, 776 
\bibitem[Steffen et al.(2006)]{Steffen} Steffen, A.~T., Strateva, I., Brandt, W.~N., Alexander, D.~M., Koekemoer, A.~M., Lehmer,  B.~D., Schneider, D.~P., \& Vignali, C.\ 2006, \aj, 131, 2826 
\bibitem[Takahashi et al. (2007)]{Takahashi}Takahashi, T., et  al.\ 2007, \pasj, 59, 35 
\bibitem[Teng et al.(2005)]{Teng05} Teng, S.~H., Wilson,  A.~S., Veilleux, S., Young, A.~J., Sanders, D.~B.,  \& Nagar, N.~M.\ 2005, \apj, 633, 664 
\bibitem[Teng et al.(2009)]{Teng} Teng, S.~H., et al.\ 2009, \apj, 691, 261 
\bibitem[Vignati et al.(1999)]{6240} Vignati, P., et al.\ 1999, \aap, 349, L57 
\bibitem[Weedman \& Houck(2008)]{Weedman} Weedman, D.~W., \& Houck, J.~R.\ 2008, \apj, 686, 127 
\bibitem[White et al.(1983)]{white} White, N.~E., Swank, J.~H., \& Holt, S.~S.\ 1983, \apj, 270, 711 
\bibitem[Wilms et al.(2000)]{wilms} Wilms, J., Allen, A., \& McCray, R.\ 2000, \apj, 542, 914 
\bibitem[Yaqoob(1997)]{yaqoob} Yaqoob, T.\ 1997, \apj, 479, 184 
\end{thebibliography}
\end{document}